\documentclass[runningheads]{llncs}
\usepackage[T1]{fontenc}
\usepackage{graphicx}
\usepackage{amsmath}
\usepackage{amsfonts}

\usepackage{listings}
\usepackage{xcolor}
\usepackage[linesnumbered,ruled,vlined]{algorithm2e}
\usepackage{hyperref}

\begin{document}
\title{From Votes to Volatility\\ Predicting the Stock Market on Election Day}

\author{Igor L.R. Azevedo\inst{1}\orcidID{0000-0001-5144-825X} \and
Toyotaro Suzumura\inst{1}\orcidID{0000-0001-6412-8386}}
\authorrunning{F. Author et al.}
\institute{The University of Tokyo, 7-chōme-3-1 Hongō, Bunkyo City, Tokyo 113-8654 \\
\email{lima-rocha-azevedo-igor@g.ecc.u-tokyo.ac.jp}, \email{suzumura@g.ecc.u-tokyo.ac.jp}}
\maketitle              % typeset the header of the contribution

\begin{abstract}
Stock market forecasting has been a topic of extensive research, aiming to provide investors with optimal stock recommendations for higher returns. In recent years, this field has gained even more attention due to the widespread adoption of deep learning models. While these models have achieved impressive accuracy in predicting stock behavior, tailoring them to specific scenarios has become increasingly important. Election Day represents one such critical scenario, characterized by intensified market volatility, as the winning candidate's policies significantly impact various economic sectors and companies. To address this challenge, we propose the \textbf{E}lection \textbf{D}ay \textbf{S}tock \textbf{M}arket \textbf{F}orecasting (\textbf{EDSMF}) Model. Our approach leverages the contextual capabilities of large language models alongside specialized agents designed to analyze the political and economic consequences of elections. By building on a state-of-the-art architecture, we demonstrate that EDSMF improves the predictive performance of SP500 during this uniquely volatile day.

\keywords{Stock Market  \and Election Day \and Stock Market Forecasting.}
\end{abstract}
\section{Introduction}
\subsection{Election Day and The Stock Market}

In democratic nations, elections play a key role in shaping government leadership, particularly presidential elections, which have a significant impact on economic policies and market dynamics. The relationship between political events and financial markets has been widely studied, focusing on how elections influence stock market behavior and volatility \cite{niederhoffer1970presidential,bialkowski2008stock,allvine1980stock}. Predicting market behavior during election periods is crucial, as outcomes can shift investor sentiment, increase volatility, and alter market efficiency, informing better decision-making and risk management for investors.

\subsection{Stock Market Forecasting}

Advances in computational power and data mining have enhanced stock price forecasting, a key topic in financial research. Traditional models like Markowitz Portfolio Optimization \cite{markowitz} introduced mean variance analysis for optimizing portfolio returns, while the Capital Asset Pricing Model (CAPM) \cite{perold2004} systematically analyzes the risk-return trade-off.

Statistical models such as ARIMA \cite{arima} and GARCH \cite{garch} have shaped stock price forecasting and risk evaluation, helping to optimize portfolio composition. Monte Carlo Simulations \cite{kroese2014montecarlo}, commonly used for pricing, also aid risk analysis and portfolio management.

Machine learning has revolutionized forecasting with models that capture complex patterns in time series data. Algorithms like Support Vector Machines (SVM) \cite{svm}, k-Nearest Neighbors (KNN) \cite{knn}, and decision tree models such as XGBoost \cite{xgboost} and CatBoost \cite{catboost} excel in predicting stock movements.

Deep learning has further advanced stock price forecasting. LSTMs \cite{lstm} capture sequential dependencies, while LSTM-RGCN \cite{lstm_rgcn} incorporates relational graph convolutional networks to model inter-stock dependencies and predict overnight price movements. To address the limitation of treating stocks as independent, STHAN-SR \cite{sthan_sr} models stock interdependence and price evolution with a spatiotemporal hypergraph attention network. ESTIMATE \cite{estimate} enhances predictions by integrating multi-order dynamics and stock behaviors using wavelet-based hypergraph convolutions for efficient message passing.

For simpler architectures, StockMixer \cite{stockmixer} uses a multi-layer perceptron (MLP)-based framework to mix indicators, time patches, and stock influences, balancing performance and efficiency. RSR \cite{rsr} combines temporal graph convolutions with relational data to model stock evolution, achieving strong returns in back-testing. HIST \cite{hist} improves forecasting accuracy by dynamically modeling stock relationships through shared information, while HATS \cite{hats} uses hierarchical attention mechanisms to aggregate relational data, enhancing predictions of individual stocks and market indices.

\subsection{Stock Market Forecasting on Election Day}

As discussed, deep learning models have significantly improved stock market prediction accuracy, making them applicable in various scenarios. Our experiments show that tailoring these models to specific conditions, such as election days with high volatility, offers substantial benefits for investors. During such periods, the inverse relationship between volatility and potential gains creates a unique profit opportunity.

To address this, we propose the \textbf{E}lection \textbf{D}ay \textbf{S}tock \textbf{M}arket \textbf{F}orecasting (\textbf{EDSMF}) Model, which enhances stock market prediction by incorporating political signals and advanced machine learning techniques. This is the first work to combine the state-of-the-art forecasting architecture, \textit{StockMixer} \cite{stockmixer}, with a novel political awareness approach. The key contributions are:

\begin{enumerate}
    \item We propose a model that enhances stock forecasting by incorporating political signals specific to each election candidate.
    \item We design a large language model (LLM)-driven agent that analyzes news articles to generate reports on candidates' economic plans and their potential impacts.
    \item We conduct extensive ablation studies to evaluate trade-offs, highlighting strengths and limitations.
\end{enumerate}

\section{Methodology}

\subsection{Overall Setting}

Building on existing works, particularly \textit{StockMixer} \cite{stockmixer}, we use normalized historical stock data with indicators like open, high, low, close, volume, daily variation, and Exponential Moving Average (EMA) as inputs. Additionally, we incorporate sector information and new \textit{political signals} to capture both market dynamics and political influences.

Our experiments focus on analyzing the impact of the 2024 United States presidential election on financial markets. The election results, along with associated news and political developments, offer a unique opportunity to study investor sentiment and market reactions. By integrating political signals, our approach aims to uncover sector-specific patterns and market-wide responses.

Unlike \textit{StockMixer}, which operates at a daily frequency, our model targets high-frequency trading, focusing on all S\&P 500 stocks during the period from 30/10/2024 to 06/11/2024 (U.S. election results). With a 1-minute dataset frequency, our model enables a more detailed analysis of stock movements. Further details about our dataset and its construction will be provided in the next section.

\subsection{Problem Definition}  

The task involves predicting the \textbf{closing price of the next data point}, which corresponds to the next minute during market hours or the next day if the market has closed. This prediction is used to calculate the 1-data-point return ratio. Given a stock market dataset consisting of $N$ stocks, where each stock $\mathcal{X}_i \in \mathbb{R}^{T \times F}$ contains historical data across $F$ indicators within a lookback window of length $T$, the return ratio $r_i^t$ at time $t$ is computed as: 
\begin{equation}
    r_i^t = \frac{p_i^t - p_i^{t-1}}{p_i^{t-1}},
\end{equation}
where $p_i^t$ is the predicted closing price.

\subsection{Input Features}  
The input dataset for the model includes a combination of traditional stock indicators, sector information, and newly introduced political signals. The following columns are used:

\begin{itemize}
     \item \textbf{open}: Opening price of the stock at the start of the given minute.
     \item \textbf{close}: Closing price of the stock at the end of the given minute.
    \item \textbf{high}: Highest price reached during the trading day.
    \item \textbf{low}: Lowest price reached during the trading day.
    \item \textbf{volume}: Total number of shares traded during the day.
    \item \textbf{daily variation}: Difference between the daily high and low prices.
    \item \textbf{EMA (Exponential Moving Average)}: A weighted moving average that gives greater importance to recent prices. The EMA at time $t$ is recursively calculated as:
    \begin{equation}
        EMA_t = \alpha \cdot p_t + (1 - \alpha) \cdot EMA_{t-1},
    \end{equation}
    where $p_t$ is the closing price at time $t$, $EMA_{t-1}$ is the EMA of the previous time step, and $\alpha$ is the smoothing factor defined as:
    \begin{equation}
        \alpha = \frac{2}{n + 1}.
    \end{equation}
    Here, $n$ is the lookback period, which is set to 45 minutes for one-minute bars.
    \item \textbf{sector}: Sector classification of the stock (e.g., technology, healthcare, energy).
\end{itemize}

\section{Extending StockMixer with Political Signals}  
While \textit{StockMixer} focuses on capturing stock-to-market, market-to-stock, and temporal interactions using multi-scale MLP-based architectures, our approach introduces \textbf{political signals} to account for the influence of national elections on financial markets. Before diving into the columns that compute this information let's first explain on how we compute and obtain such information.

\subsection{Candidate Impact}

\subsubsection{LLM-Agent Framework for Political Signal Generation}\label{sec:agents}  

To generate the political signals, we leverage a Large Language Model (LLM)-\textit{agents} framework consisting of four specialized agents:

\begin{enumerate}
    \item \textbf{News Analyst}  
    \begin{itemize}
        \item \textit{Role}: Analyzes recent political news and developments for both candidates.  
        \item \textit{Output}: Summarized campaign dynamics and key policy announcements.  
    \end{itemize}

    \item \textbf{Candidate Policy Analyst}  
    \begin{itemize}
        \item \textit{Role}: Compares and contrasts the candidates' economic policies, including taxation, regulation, and trade.  
        \item \textit{Output}: Detailed markdown comparison of policy stances and their potential economic impact.  
    \end{itemize}

    \item \textbf{Market Analyst}  
    \begin{itemize}
        \item \textit{Role}: Evaluates how different sectors of the S\&P 500 might be affected by each candidate's policies.  
        \item \textit{Output}: Sector-specific analysis with clear rankings of sectors likely to benefit or face challenges under each scenario.  
    \end{itemize}

    \item \textbf{Synthesis Analyst}  
    \begin{itemize}
        \item \textit{Role}: Integrates outputs from all previous agents to connect political developments to specific market outcomes.  
        \item \textit{Output}: A comprehensive markdown report highlighting impacted sectors, expected outcomes, and investment recommendations under each candidate's potential victory.  
    \end{itemize}
\end{enumerate}

To create the political signals, we analyzed over 90,000 news articles from various U.S. news sources, published between \textit{2024-01-01} and \textit{2024-11-06}, providing a comprehensive overview of the policies and campaign dynamics of the candidates. Additionally, we incorporated official campaign documents \footnote{\url{https://kamalaharris.com/wp-content/uploads/2024/09/Policy\_Book\_Economic-Opportunity.pdf}, \url{https://rncplatform.donaldjtrump.com/}} to enhance the model's understanding of their economic proposals. This multi-source approach ensures a robust and nuanced dataset for generating the political signals.

\begin{table}[ht]
\centering
\caption{Sector Impacts by Candidate}
\label{tab:sector_impact}
\begin{tabular}{|l|c|c|}
\hline
\textbf{Sector}                & \textbf{Candidate 1 Impact} & \textbf{Candidate 2 Impact} \\ \hline
Energy                         & Positive                   & Negative                   \\ \hline
Industrials                    & Positive                   & Negative                   \\ \hline
Financials                     & Positive                   & Negative                   \\ \hline
Materials                      & Positive                   & Negative                   \\ \hline
Consumer Discretionary         & Positive                   & Negative                   \\ \hline
Real Estate                    & Positive                   & Negative                   \\ \hline
Utilities                      & Positive                   & Negative                   \\ \hline
Communication Services         & Positive                   & Negative                   \\ \hline
Information Technology         & Negative                   & Positive                   \\ \hline
Health Care                    & Negative                   & Positive                   \\ \hline
Consumer Staples               & Negative                   & Positive                   \\ \hline
\end{tabular}
\end{table}

\subsubsection{Formulation}

The candidate impact variables, \textit{candidate\_impact\_1} and \textit{candidate\_impact\_2}, quantify the expected market impact of each candidate based on their proposed economic policies and campaign momentum. These features are designed to capture the financial market's perception of how each candidate's potential leadership might influence individual stocks and various economic sectors. Table \ref{tab:sector_impact} summarizes the sectoral impacts as determined by our agents, described in Section \ref{sec:agents}.

For a stock $i$ at time $t$, the \textit{candidate impact} is computed as:
\begin{equation}
    \text{candidate\_impact}_{c,i,t} = \sum_{s=1}^{S} \mathbb{I}_{c,s} \cdot \mathbb{I}_{s,i},
\end{equation}
where:
\begin{itemize}
    \item $c \in \{1, 2\}$ represents the candidate index (Candidate 1 or Candidate 2).
    \item $S$ is the total number of economic sectors.
    \item $\mathbb{I}_{c,s}$ is an indicator function assigning a value of $1$ if the candidate has a positive impact on sector $s$, and $-1$ if the impact is negative.
    \item $\mathbb{I}_{s,i}$ is an indicator function that assigns a value of $1$ if stock $i$ belongs to sector $s$, and $0$ otherwise.
\end{itemize}

This formulation effectively assigns a score of $1$ to stocks in sectors with a positive impact under a candidate's policies, and $-1$ to stocks in sectors with a negative impact. The approach ensures that sector-level policy impacts are accurately mapped to individual stocks, enabling the model to capture nuanced market reactions to each candidate's proposed policies.

\subsection{Candidate Context}

The \textbf{candidate\_context} variable encodes the prevailing political scenario by assigning each data point to one of the two candidates contesting the election. This variable ensures the model can account for candidate-specific market influences during training and validation. The candidate context is implemented in two distinct strategies: \textit{random assignment} for training and validation, and a consistent \textit{ensemble strategy} for the test phase.

\subsubsection{Random Assignment Strategy}

During the training and validation phases, the \textit{candidate\_context} variable is randomly assigned to represent one of the two candidates contesting the election. This variable allows the model to learn candidate-specific influences on stock market behavior.

\begin{algorithm}[H]
\SetAlgoLined
\KwIn{Dataset containing stocks and their associated timestamps, $D$}
\KwOut{Dataset with assigned candidate contexts, $D_{context}$}

\ForEach{row in $D$}{
    \uIf{$\text{date\_day} < \text{Election\_Day}$}{
        Randomly assign \textit{candidate\_context} as:
        \begin{equation}
            \text{candidate\_context} = 
            \begin{cases}
                1 & \text{with probability } 0.5, \\
                2 & \text{with probability } 0.5
            \end{cases}
        \end{equation}
    }
    \Else{
        Assign \textit{candidate\_context} = 1 (representing the winning candidate)
    }
}
\Return{$D_{context}$}
\caption{Random Assignment of Candidate Context}
\label{alg:candidate-context}
\end{algorithm}

For a data point at time $t$, the random assignment is defined as:
\begin{equation}
    \text{candidate\_context}_t = 
    \begin{cases} 
      1 & \text{if Candidate 1 is assigned}, \\
      2 & \text{if Candidate 2 is assigned}.
    \end{cases}
\end{equation}

The assignment logic is implemented using the procedure shown in algorithm \ref{alg:candidate-context}. The random assignment strategy ensures that the model can generalize well by learning the relationship between stock behavior and the influence of both candidates during the training process. By consistently setting the \textit{candidate\_context} during testing to reflect the actual election outcome, the model's predictions remain realistic and aligned with real-world scenarios.

\subsubsection{Ensemble Strategy}

The \textit{ensemble strategy} provides a robust mechanism for incorporating multiple candidate contexts into the stock prediction process. Instead of using a single random assignment, as in the random strategy, the ensemble strategy trains two separate models with distinct candidate contexts and combines their predictions during validation and testing. This approach ensures that the model captures the market's reaction to both candidates while maintaining flexibility in adapting to the real-world scenario.

\paragraph{Training and Validation Phases}

During training and validation, two separate models, \textit{Model A} and \textit{Model B}, are trained with fixed candidate contexts:
\begin{itemize}
    \item \textbf{Model A:} Candidate context is set to \textbf{1}.
    \item \textbf{Model B:} Candidate context is set to \textbf{2}.
\end{itemize}

For a stock $i$ at time $t$, the \textit{candidate\_context} for training and validation is defined as:
\begin{equation}
    \text{candidate\_context}_{t} =
    \begin{cases}
        1, & \text{for Model A (training and validation)} \\
        2, & \text{for Model B (training and validation)}
    \end{cases}
\end{equation}

This dual-model setup enables the system to separately learn how the stock market would respond to each candidate's policies.

\paragraph{Testing Phase}

During testing, the candidate context is consistently set to the actual winner of the election (Candidate 1). For both models, the candidate context is defined as:
\begin{equation}
    \text{candidate\_context}_{t} = 1, \quad \forall t \in \text{test period}.
\end{equation}

\paragraph{Ensemble Prediction}

The ensemble strategy integrates the predictions from both models during the validation and test phases. The final prediction is computed as a weighted average of predictions from both models:
\begin{equation}
    \text{prediction}_{\text{ensemble}, t} = w_1 \cdot \text{prediction}_{A,t} + w_2 \cdot \text{prediction}_{B,t},
\end{equation}
where:
\begin{itemize}
    \item $w_1$ and $w_2$ are the ensemble weights for \textit{Model A} and \textit{Model B}, respectively.
    \item $\text{prediction}_{A,t}$ and $\text{prediction}_{B,t}$ are the predictions from Model A and Model B at time $t$.
\end{itemize}

The ensemble weights reflect the importance of each candidate in influencing the market during the test period. For instance, in the provided code, the weights are set as:
\begin{equation}
    w_1 = 0.2, \quad w_2 = 0.8.
\end{equation}

This weighting scheme prioritizes the influence of Candidate 2 while still considering the impact of Candidate 1.

\begin{algorithm}[H]
\SetAlgoLined
\KwIn{Training dataset $D_{\text{train}}$, Validation dataset $D_{\text{valid}}$, Test dataset $D_{\text{test}}$}
\KwOut{Ensemble predictions for test dataset, $\text{predictions}_{\text{ensemble}}$}

\ForEach{row in $D_{\text{train}} \cup D_{\text{valid}}$}{
    Assign \textit{candidate\_context} = 1 for Model A\;
    Assign \textit{candidate\_context} = 2 for Model B\;
}

\ForEach{row in $D_{\text{test}}$}{
    Assign \textit{candidate\_context} = 1 (reflecting the winning candidate)\;
}

Train \textbf{Model A} on $D_{\text{train}}$ with \textit{candidate\_context} = 1\;
Train \textbf{Model B} on $D_{\text{train}}$ with \textit{candidate\_context} = 2\;

Validate both models on $D_{\text{valid}}$ and compute predictions $\text{prediction}_{A}$ and $\text{prediction}_{B}$\;

Combine predictions during testing as:
\begin{equation}
    \text{prediction}_{\text{ensemble}} = w_1 \cdot \text{prediction}_{A} + w_2 \cdot \text{prediction}_{B}.
\end{equation}

\Return{$\text{prediction}_{\text{ensemble}}$}
\caption{Ensemble Strategy for Candidate Context}
\label{alg:ensemble-context}
\end{algorithm}

\subsection{Integration into StockMixer}

By extending the StockMixer architecture to include \textit{candidate\_impact\_1}, \textit{candidate\_impact\_2}, and \textit{candidate\_context} as additional input features, we enable the model to account for political signals during election periods. These variables allow the model to: (1) Capture candidate-specific effects on stock behavior, (2) Align predictions with real-world election outcomes, (3) Enhance forecasting accuracy during high-volatility periods. This integration represents a novel approach to political-aware stock market forecasting.

\subsection{Final Expression}  
The extended framework predicts the closing price $p_i^t$ and the corresponding 1-data-point return ratio $r_i^t$, where the input data $\mathcal{X} \in \mathbb{R}^{N \times T \times (F+3)}$ includes the additional political features:

\begin{equation}
    \mathcal{X} \in \mathbb{R}^{N \times T \times (F+3)} \xrightarrow{\theta} p \in \mathbb{R}^{N \times 1} \rightarrow r \in \mathbb{R}^{N \times 1}.
\end{equation}

\subsection{Loss Function}

Following \cite{stockmixer} use the \textbf{1-data-point (1 minute) return ratio} of a stock as the ground truth rather than the normalized price. This approach combines a pointwise regression and pairwise ranking-aware loss to minimize the MSE between the predicted and actual return ratios while maintaining the relative order of top-ranked stocks with higher expected returns for investment as:

\begin{equation} \label{eq:loss_function}
    L = L_{\text{MSE}} + \alpha \sum_{i=1}^{N} \sum_{j=1}^{N} \max(0, -( \hat{r}_i^t - \hat{r}_j^t )(r_i^t - r_j^t)),
\end{equation}

\noindent where $\hat{r}^t$ and $r^t$ are the predicted and actual ranking scores, respectively, and $\alpha$ is a weighting parameter.

\section{Experiments}

\subsection{Experimental Setup}

\subsubsection{Datasets}

We evaluated our model on the United States 2024 presidential election using high-frequency stock market data sampled every minute between 2024-10-30 and 2024-11-06. The dataset was divided into three splits as follows:

\begin{itemize}
    \item \textbf{Training:} From 2024-10-30 (09:30) to 2024-11-04 (15:59), totaling 1,559 data points.
    \item \textbf{Validation:} From 2024-11-05 (09:30) to 2024-11-05 (15:59), totaling 389 data points.
    \item \textbf{Testing (US Election Day):} From 2024-11-06 (09:30) to 2024-11-06 (15:59), totaling 389 data points.
\end{itemize}

\subsubsection{Infrastructure}

The implementation of the deep learning model is based on the \textit{StockMixer} \cite{stockmixer} model, utilizing PyTorch\footnote{\url{https://pytorch.org/}} for the core computations. For the LLM-agent model, we employed CrewAI\footnote{\url{https://www.crewai.com/}}. The source codes for both components are publicly available: \href{https://github.com/toyolabo/EDSMF}{Code Repository}.

To ensure a fair comparison, all samples were generated by sliding a 16-day lookback window across minute level data points. The experiments were conducted with a fixed loss factor $\alpha = 0.1$, a learning rate of $10^{-3}$ and 100 epochs. All computations were performed on a server equipped with one NVIDIA A100 GPU.

\subsubsection{Evaluation Metrics}

To comprehensively evaluate the performance of the proposed model, we employed the same four metrics reported by \cite{stockmixer}, which are widely used in financial forecasting.

\begin{itemize}
    \item \textbf{Information Coefficient (IC):} Measures the Pearson correlation coefficient between predicted and actual results. It evaluates how well the predictions align with the ground truth.
    \item \textbf{Rank Information Coefficient (RIC):} Computes the Spearman correlation coefficient for the ranked profit potentials of stocks. This metric assesses the model's stock selection ability and is closely tied to rank loss.
    \item \textbf{Precision@N:} Evaluates the precision of the top $N$ predictions.
    \item \textbf{Sharpe Ratio (SR):} Balances return and risk by calculating the average return per unit of volatility relative to the risk-free rate:
    \begin{equation}
        SR = \frac{R_t - R_f}{\theta},
    \end{equation}
    where $R_t$ is the return, $R_f$ is the risk-free rate, and $\theta$ is the standard deviation of the returns.
\end{itemize}

\subsection{Results}

To ensure fair and consistent results, each model was trained and evaluated at least three times. The average performance was calculated for each run, and the checkpoint with the lowest loss was selected, as defined by the loss function in Equation \ref{eq:loss_function}. The evaluation considered multiple metrics, including IC, RIC, Prec@N, and SR, to assess both the predictive accuracy and risk-adjusted returns.

Table \ref{tab:best-ensemble-metrics} presents a comparison of the best ensemble configurations for each metric, alongside the baseline model (\textit{StockMixer}) and a random assignment strategy. The model configuration `20-Candidate-1, 80-Candidate-2` achieved the highest RIC (0.2306) and competitive performance across other metrics. The `40-Candidate-1, 60-Candidate-2` configuration excelled in Sharpe Ratio (1.8163) and IC (0.0898), demonstrating the robustness of ensemble strategies in capturing candidate-specific influences.

\begin{table}[h!]
\centering
\caption{Comparison of Metrics: Best Ensemble for Each Metric, Random, and Baseline Models}
\begin{tabular}{lcccc}
\hline
\textbf{Model} & \textbf{Prec@10} & \textbf{IC} & \textbf{RIC} & \textbf{SR} \\
\hline
\textbf{20-Candidate-1, 80-Candidate-2} & \textbf{0.5829} & 0.0770 & \textbf{0.2306} & 1.7660 \\
\textbf{40-Candidate-1, 60-Candidate-2} & 0.5670 & \textbf{0.0898} & 0.2209 & \textbf{1.8163} \\
\textbf{80-Candidate-1, 20-Candidate-2} & \underline{0.5783} & 0.0794 & \underline{0.2301} & 1.7680 \\
\textbf{60-Candidate-1, 40-Candidate-2} & 0.5738 & 0.0797 & 0.2163 & \underline{1.7714} \\
\textbf{Random Assignment} & 0.5693 & 0.0797 & 0.2000 & 1.7710 \\
\textbf{StockMixer (Baseline)} & 0.5694 & \underline{0.0860} & 0.2290 & 1.7704 \\
\hline
\end{tabular}
\label{tab:best-ensemble-metrics}
\end{table}

\begin{figure}[h!]
\centering
\includegraphics[width=\textwidth]{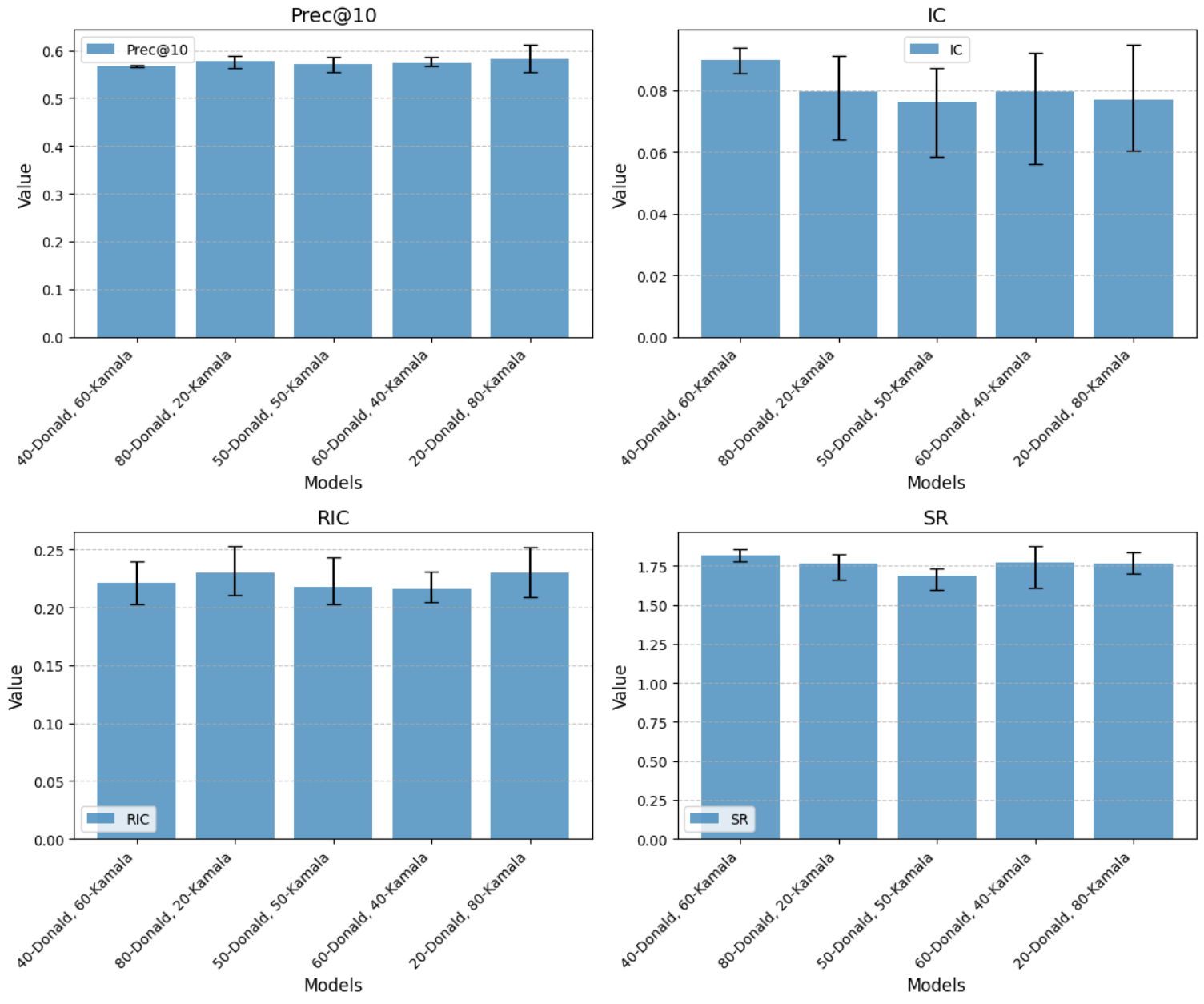}
\caption{Comparison of performance metrics (`Prec@10`, `IC`, `RIC`, `SR`) for various ensemble configurations. Each bar chart represents a single metric, highlighting the contribution of different weights assigned to candidates 1 and 2. Models are labeled as `weight1-weight2`, where `weight1` represents the percentage for candidate 1, and `weight2` represents the percentage for candidate 2.}
\label{fig:ensemble-metrics}
\end{figure}

\section{Ablation Studies}

To further investigate the contribution of candidate-specific weights to model performance, we conducted a series of ablation studies with varying candidate impact ratios. Figure \ref{fig:ensemble-metrics} visualizes the performance of different ensemble configurations (`Prec@10`, `IC`, `RIC`, and `SR`) by assigning different weight distributions to candidates. The results indicate:

\begin{itemize}
    \item \textbf{Prec@10:} Higher candidate-2 weights generally improved precision, reflecting a stronger alignment between candidate-2 policies and top-performing stocks.
    \item \textbf{IC and RIC:} Configurations with balanced or slightly candidate-1-dominant weights (e.g., `40-Candidate-1, 60-Candidate-2`) provided the best correlation with actual stock performance, validating the importance of including both candidates' influences.
    \item \textbf{Sharpe Ratio:} The configuration `40-Candidate-1, 60-Candidate-2` demonstrated the highest Sharpe Ratio, highlighting its superior risk-adjusted returns.
\end{itemize}

\section{Conclusion}

This study explores the intersection of political events and stock market forecasting using high-frequency trading data from the 2024 U.S. presidential election. We propose the \textbf{E}lection \textbf{D}ay \textbf{S}tock \textbf{M}arket \textbf{F}orecasting (\textbf{EDSMF}) model, which integrates advanced methodologies with political signals generated by a large language model (LLM)-driven agent framework. This model captures candidate-specific impacts and contexts, offering a detailed understanding of stock behavior during periods of high volatility.

The results show that the EDSMF model improves predictive accuracy and risk-adjusted returns. Ablation studies highlight the importance of balancing candidate influence weights and the robustness of ensemble strategies. The inclusion of political signals enabled the model to outperform the baseline \textit{StockMixer} framework in key metrics, such as Information Coefficient (IC), Rank Information Coefficient (RIC), Precision@10, and Sharpe Ratio (SR).

In conclusion, the EDSMF model bridges political awareness and financial modeling, emphasizing the importance of scenario-specific forecasting architectures. We plan as future research to expand this framework by incorporating additional political and macroeconomic variables and exploring the framework with other countries election periods.


\begin{thebibliography}{8}

\bibitem{niederhoffer1970presidential}
Niederhoffer, V., Gibbs, S., Bullock, J.: Presidential elections and the stock market. \textit{Financial Analysts Journal} \textbf{26}(2), 111--113 (1970). \url{http://www.jstor.org/stable/4470664}

\bibitem{bialkowski2008stock}
Białkowski, J., Gottschalk, K., Wisniewski, T. P.: Stock market volatility around national elections. \textit{Journal of Banking \& Finance} \textbf{32}(9), 1941--1953 (2008). \doi{10.1016/j.jbankfin.2007.12.021}. \url{https://www.sciencedirect.com/science/article/pii/S0378426607004219}

\bibitem{allvine1980stock}
Allvine, F. C., O'Neill, D. E.: Stock market returns and the presidential election cycle: Implications for market efficiency. \textit{Financial Analysts Journal} \textbf{36}(5), 49--56 (1980). \doi{10.2469/faj.v36.n5.49}. \url{https://doi.org/10.2469/faj.v36.n5.49}


\bibitem{markowitz}
Markowitz, H.: Portfolio selection. \textit{The Journal of Finance} \textbf{7}(1), 77--91 (1952). \url{http://www.jstor.org/stable/2975974}

\bibitem{perold2004}
Perold, A. F.: The capital asset pricing model. \textit{The Journal of Economic Perspectives} \textbf{18}(3), 3--24 (2004). \url{http://www.jstor.org/stable/3216804}

\bibitem{arima}
Piccolo, D.: A distance measure for classifying ARIMA models. \textit{Journal of Time Series Analysis} \textbf{11}(2), 153--164 (1990). \doi{10.1111/j.1467-9892.1990.tb00048.x}. \url{https://onlinelibrary.wiley.com/doi/abs/10.1111/j.1467-9892.1990.tb00048.x}


\bibitem{garch}
Bollerslev, T.: Generalized autoregressive conditional heteroskedasticity. \textit{Journal of Econometrics} \textbf{31}(3), 307--327 (1986). \doi{10.1016/0304-4076(86)90063-1}. \url{https://www.sciencedirect.com/science/article/pii/0304407686900631}


\bibitem{kroese2014montecarlo}
Kroese, D. P., Brereton, T., Taimre, T., Botev, Z. I.: Why the Monte Carlo method is so important today. \textit{WIREs Computational Statistics} \textbf{6}(6), 386--392 (2014). \doi{10.1002/wics.1314}. \url{https://wires.onlinelibrary.wiley.com/doi/abs/10.1002/wics.1314}


\bibitem{svm}
Hearst, M. A., Dumais, S. T., Osuna, E., Platt, J., Scholkopf, B.: Support vector machines. \textit{IEEE Intelligent Systems and their Applications} \textbf{13}(4), 18--28 (1998). \doi{10.1109/5254.708428}

\bibitem{knn}
Peterson, L. E.: K-nearest neighbor. \textit{Scholarpedia} \textbf{4}(2), 1883 (2009). \doi{10.4249/scholarpedia.1883}


\bibitem{xgboost}
Chen, T., Guestrin, C.: XGBoost: A scalable tree boosting system. In: \textit{Proceedings of the 22nd ACM SIGKDD International Conference on Knowledge Discovery and Data Mining}, pp. 785--794. ACM, 2016. \doi{10.1145/2939672.2939785}. \url{http://dx.doi.org/10.1145/2939672.2939785}


\bibitem{catboost}
Prokhorenkova, L., Gusev, G., Vorobev, A., Dorogush, A. V., Gulin, A.: CatBoost: unbiased boosting with categorical features. arXiv preprint arXiv:1706.09516 (2019). \url{https://arxiv.org/abs/1706.09516}

\bibitem{lstm}
Graves, A.: Long short-term memory. In: \textit{Supervised Sequence Labelling with Recurrent Neural Networks}, pp. 37--45. Springer, Berlin, Heidelberg (2012). \doi{10.1007/978-3-642-24797-2_4}. \url{https://doi.org/10.1007/978-3-642-24797-2\_4}

\bibitem{lstm_finance}
Fischer, T., Krauss, C.: Deep learning with long short-term memory networks for financial market predictions. \textit{European Journal of Operational Research} \textbf{270}(2), 654--669 (2018). \doi{10.1016/j.ejor.2017.11.054}. \url{https://www.sciencedirect.com/science/article/pii/S0377221717310652}

\bibitem{lstm_rgcn}
Li, W., Bao, R., Harimoto, K., Chen, D., Xu, J., Su, Q.: Modeling the stock relation with graph network for overnight stock movement prediction. In: \textit{Proceedings of the Twenty-Ninth International Conference on International Joint Conferences on Artificial Intelligence}, pp. 4541--4547 (2021)


\bibitem{sthan_sr}
Sawhney, R., Agarwal, S., Wadhwa, A., Derr, T., Shah, R. R.: Stock selection via spatiotemporal hypergraph attention network: A learning to rank approach. \textit{Proceedings of the AAAI Conference on Artificial Intelligence} \textbf{35}(1), 497--504 (2021). \doi{10.1609/aaai.v35i1.16127}. \url{https://ojs.aaai.org/index.php/AAAI/article/view/16127}

\bibitem{estimate}
Huynh, T. T., Nguyen, M. H., Nguyen, T. T., Nguyen, P. L., Weidlich, M., Nguyen, Q. V. H., Aberer, K.: Efficient integration of multi-order dynamics and internal dynamics in stock movement prediction. arXiv preprint arXiv:2211.07400 (2022). \url{https://arxiv.org/abs/2211.07400}

\bibitem{stockmixer}
Fan, J., Shen, Y.: StockMixer: A simple yet strong MLP-based architecture for stock price forecasting. \textit{Proceedings of the AAAI Conference on Artificial Intelligence} \textbf{38}(8), 8389--8397 (2024). \doi{10.1609/aaai.v38i8.28681}. \url{https://ojs.aaai.org/index.php/AAAI/article/view/28681}

\bibitem{rsr}
Feng, F., He, X., Wang, X., Luo, C., Liu, Y., Chua, T.-S.: Temporal relational ranking for stock prediction. \textit{ACM Transactions on Information Systems} \textbf{37}(2), 1--30 (2019). \doi{10.1145/3309547}. \url{http://dx.doi.org/10.1145/3309547}

\bibitem{hist}
Xu, W., Liu, W., Wang, L., Xia, Y., Bian, J., Yin, J., Liu, T.-Y.: HIST: A graph-based framework for stock trend forecasting via mining concept-oriented shared information. arXiv preprint arXiv:2110.13716 (2022). \url{https://arxiv.org/abs/2110.13716}

\bibitem{hats}
Kim, R., So, C. H., Jeong, M., Lee, S., Kim, J., Kang, J.: HATS: A hierarchical graph attention network for stock movement prediction. arXiv preprint arXiv:1908.07999 (2019). \url{https://arxiv.org/abs/1908.07999}

% -----------------------

\end{thebibliography}
%

\end{document}